\documentclass[fleqn,a4paper,12pt,twoside]{elsart}
        
\usepackage[figuresright]{rotating}
\usepackage{graphicx}    
\usepackage{natbib}
\newcommand{\be}{\begin{equation}}
\newcommand{\cm}{{~\:\rm cm}}
\newcommand{\ee}{\end{equation}}
\newcommand{\erg}{{~\:\rm erg}}

\newcommand{\msun}{~\:{\mathcal M}_\odot}
\newcommand{\pc}{{~\:\rm pc}}
\newcommand{\yr}{{~\:\rm yr}}

\begin{document}
\begin{frontmatter}

\title{Binary Black Holes at the Core of Galaxy Clusters}

\author{Fabio Pizzolato}
\and
\author{Noam Soker}
\address{Department of Physics\\
    Technion, Israel Institute of Technology\\ 
    Technion City, 32000 Haifa, Israel\\ }  


\runtitle{Binary Black Holes}
\runauthor{Fabio Pizzolato \& Noam Soker}


\begin{abstract}
In this paper we push forward and exploit an  analogy between the morphologies
of the X-ray cavities observed in some galaxy
clusters, and the optically deficient point-symmetric bubbles  
occurring in some planetary nebulae (PNe). 
Point-symmetric PNe are thought to be shaped by stellar binary
interactions; namely, the presence of a
companion to the PN's progenitor star is required. 
We suggest that similar point-symmetric structures in 
the X-ray cavities of galaxy clusters might be
associated with the presence of massive binary black holes. 
A systematic cataloguing of  high-resolution images of the  
diffuse X-ray emission at the core of galaxy clusters might  
contribute to individuate massive binary black holes.
\end{abstract}

\end{frontmatter}


\section{Introduction}

\label{s:intro}

In our   hierarchical  Universe, where  
small structures merge into larger ones, massive 
binary black holes (i.e. with $M=10^8-10^9\msun$) are believed to be fairly 
common.
Yet, up to now there is little compelling  evidence of their existence  
\citep{Kom03b, Mer04}. The  strongest candidate for a massive 
binary black  hole 
(MBBH) are  two X-ray active AGNs at the core of the nearby ($z=0.02$) galaxy 
NGC~6240 observed with {\sl Chandra}  \citep{Kom03a}. 
These AGNs are only 1.5~arcsec apart, and this hints that  the possibility 
for a ``direct'' detection of double AGNs is limited to very nearby objects. 
Other pieces of evidence are more indirect (\citealp{Beg80},  see also
\citealp{Kom03b} for a  recent review). 

The history of a massive
binary black hole may be summarised in the following steps 
\citep{Beg80,  Mer04, Roo88}:
$1)$~after two galaxies merge, the two massive black holes in the former
galaxies' nuclei sink into the gravitational well on account of the dynamical 
friction, and they eventually form a MBBH; $2)$~the binary orbit slowly
decays, since the interactions with field stars 
extract  energy and angular  momentum from the binary system;
$3)$~if the binary separation decreases
enough, gravitational radiation takes over and the   
binary rapidly coalesces.

The centre  of cool-core galaxy clusters is almost always occupied by a 
massive cD galaxy, formed by  the mergers of several smaller 
galaxies \citep{Dub98}. cD galaxies, therefore, seem the ideal 
environment to  look for MBBHs.

Step 1 is quite fast, while the passage from phase~2 to phase~3 may take a 
very long time \citep{Mer04}. Therefore, many MBBH are expected to lie in the
intermediate phase~2, where their separation is still too large to 
allow a significant gravitational radiation emission, but where the
mutual gravitational interaction is strong enough to spark interesting 
phenomena.
 
In this paper we discuss the possibility to detect MBBH at the core of galaxy
clusters from a morphological analysis of their X-ray emission. Our 
method is based on an analogy between the morphology of  the
diffuse X-ray emission at the core of some galaxy clusters, and similar
structures observed in the optical emission of some planetary nebulae
(PNe) \citep{Sok03b,Sok03,Sok04}.
This similarity suggests that a common physical mechanism lies 
at the origin of these structures. We argue  that this common mechanism is the 
gravitational action  of a binary system.

\section{Point-Symmetric Bubbles in Planetary Nebulae}

\begin{figure}[htb]
\includegraphics[width =75mm, angle = -90]{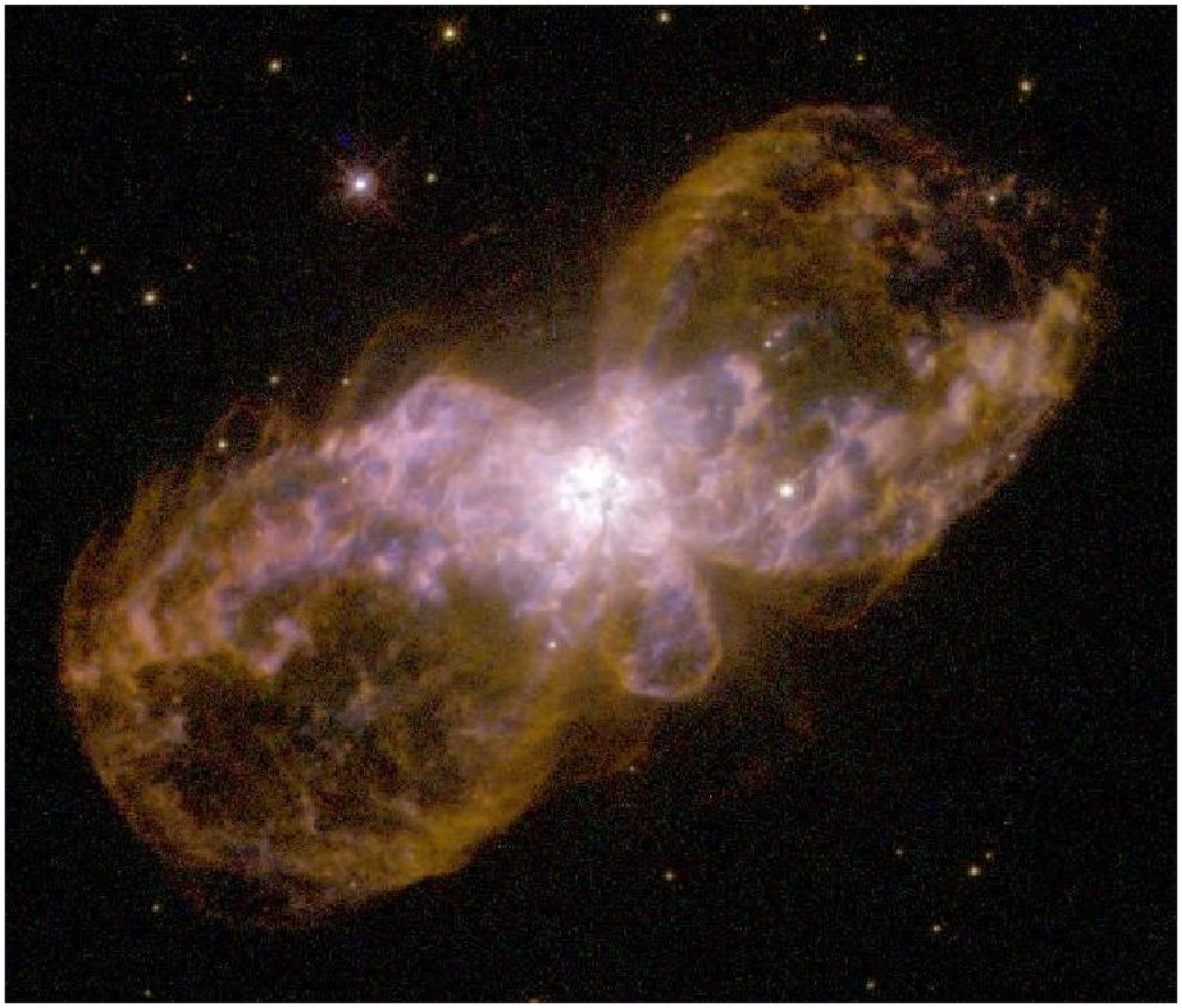}
\hspace{\fill}
\includegraphics[width =75mm, angle = -90]{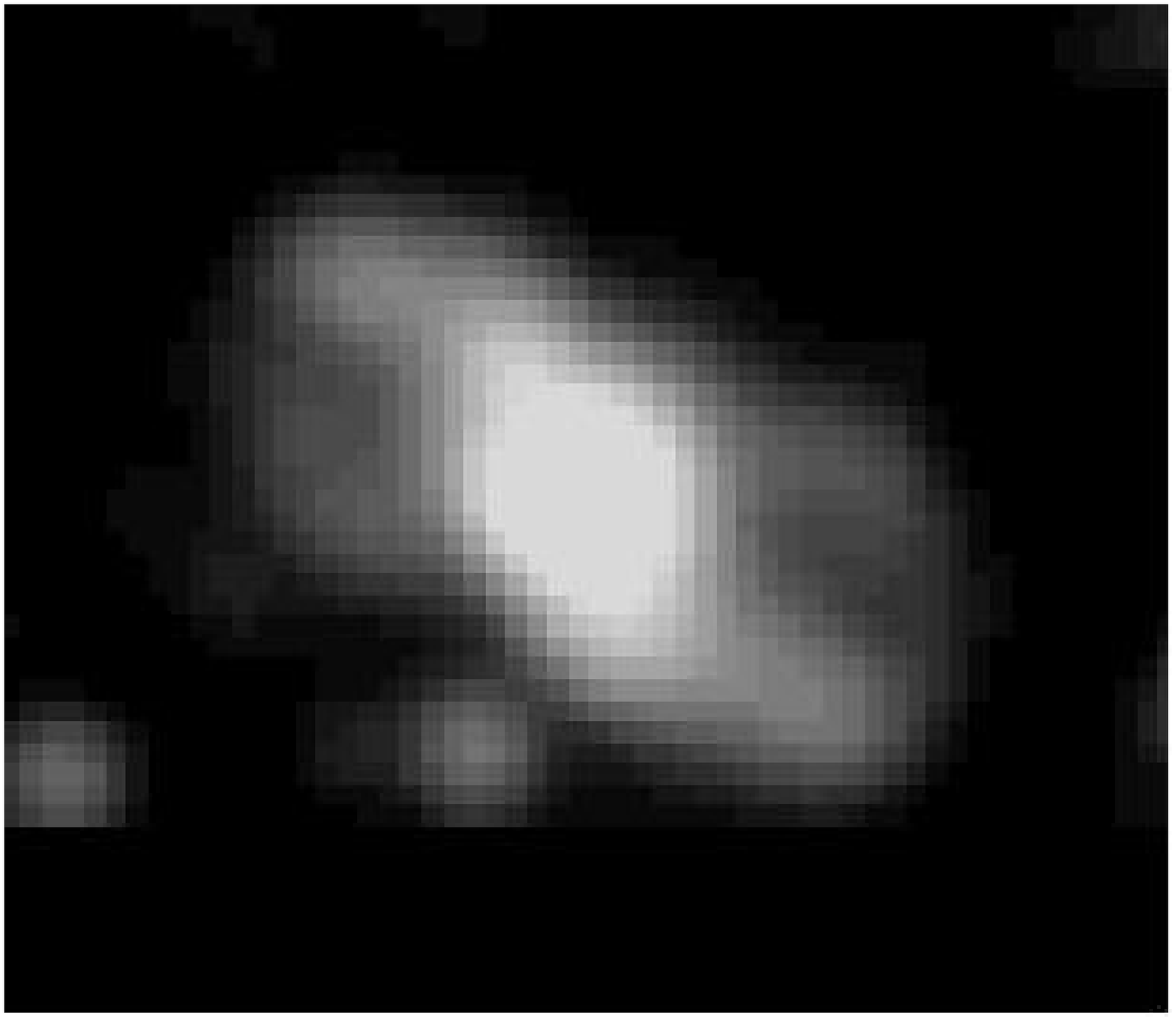}
\caption{\label{f:hb5} The planetary nebula Hb~5.
Left panel: a HST high-resolution image \citep{Ter00}; image by kind 
permission of A.~R.~Hajian, see  http://ad.usno.navy.mil/pne/gallery.html.
Right panel: a low-resolution image of the same object
from the catalogue of \citet{Sch92}, and the original resolution was 
degraded by Gaussian smoothing (R.~Corradi, private communication).
The  edge-to-edge angular distance along the major axis  is 60",
corresponding to a linear scale of about $0.4\pc$ at the nebula's distance.
}

\hspace{\fill}
\begin{minipage}[t]{70mm}
\includegraphics[width = 70mm, angle = 0]{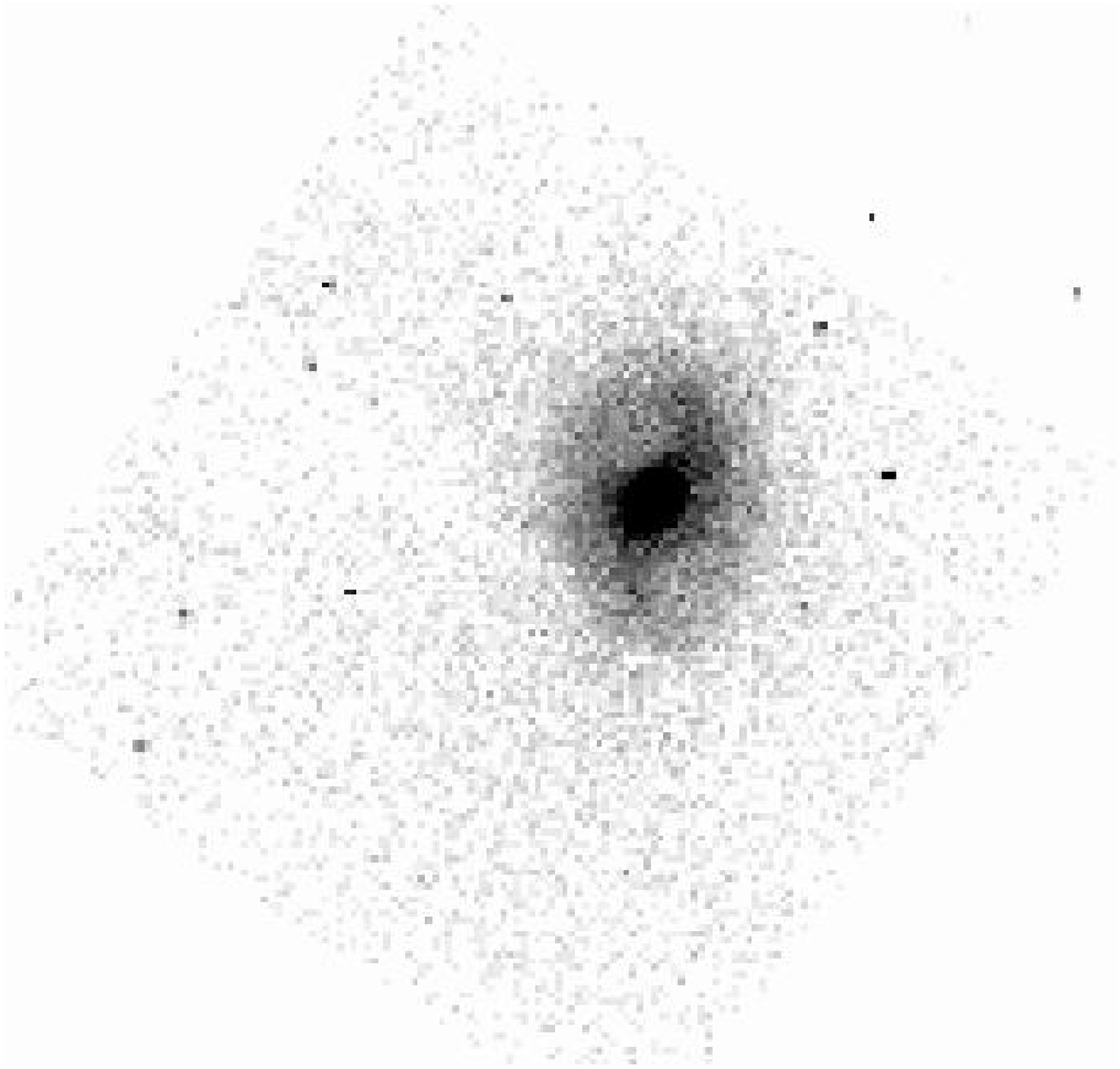}
\caption{\label{f:ms} A {\sl Chandra} X-ray image of the galaxy cluster 
MS~0735.6+7421 exhibits a  point-symmetric structure.
The cavities near the ``arms'' have a diameter of  about 250~kpc 
\citep{McN05}.}
\end{minipage}
\end{figure}

Solar-sized stars loose their outer envelope during the Asymptotic Giant Branch
(AGB) phase, after they quit the main sequence. The lost shell is ionised by 
the leftover central white dwarf, and shines as a PN.
Actually, only $10\%$~ of all PNe are spherically symmetric \citep{Sok92};
most of them are axisymmetric, or even completely asymmetric, with
a wide spectrum of morphologies. Here we focus on bipolar PNe,
which exhibit two lobes and an equatorial  waist 
between them. A prototype of this kind of PNe is  Hb~5, shown in 
Figure~\ref{f:hb5}. {\em Point-symmetry} is another feature of 
this nebula.
As in geometry, a structure is said to be {\em point-symmetric} if, 
to each substructure,
a similar substructure corresponds on the opposite side with respect  to
the centre. In general, point symmetry does not imply neither mirror 
symmetry nor   axial symmetry. 


A pair of 
optically faint lobes is also apparent in this object, as well as
in others PNe. 
The inner regions of the lobes are optically faint on account of their low 
density, which is about 2-3 orders of magnitude  below  the surroundings 
\citep{Sok03}.

The formation of bipolar PNe and point-symmetric structures can 
be explained by the influence of a binary companion  \citep{Sok04}. 
The AGB wind from the primary star forms an accretion disc around a 
secondary  compact star; in turn, this accretion disc  
launches   jets  and/or collimated  fast winds into the PN shell.
If the jet orientation remains constant in time, the PN
should possess a pure axisymmetrical structure. The bright pattern is
different if the jets precess, however. 
During the precession motion the jets expand in a point-symmetric 
geometry centred on the PN core, hence inflating 
point-symmetric bubble pairs.
 
What is the cause of this precession? If the jet is fired by an accretion 
disc, the jet's precession is  a direct consequence of the disc's own 
precession. Different  mechanisms may be able to drive it.

The first one is the action of a companion star, which makes the disc 
precess at a rate \citep{Kat97} 
\be
\omega_{\rm prec} \simeq 0.4 \left(\frac{G\, M}{a^3}\right)^{1/2} \; 
\left(\frac{a_d}{a}\right)^{3/2} \;  \frac{q}{(1+q)^{1/2}}\; \cos \vartheta,
\ee 
where $q=M_2/M_1$, $M_1$ is the mass of the accreting object, $M_2$ is 
the mass of the mass-loosing component, $M= M_1+M_2$ is the mass of the 
binary system, $a$ is the  separation  between the components 
(assuming a circular orbit), $a_d$  is the disc's radius, and
$\vartheta$ is the tilt  angle between the disc and  the orbital plane. 
Normalising to values typical for a stellar binary system, the precession 
period  $\tau_{\rm prec}=2\pi/\omega_{\rm prec}$ is 
\be
\tau_{\rm prec} \simeq 10^3  \; \yr \; 
\left(\frac{M}{2 \msun}\right)^{1/2}
\left(\frac{a}{10^{14}\cm}\right)^3
\left(\frac{a_d}{10^{13}\cm}\right)^{-3/2}\; 
\frac{(1+q)^{1/2}}{q\: \cos\vartheta}.
\ee

A second possible scenario is that the PN's progenitor is a single star.
In this case, the mechanism driving the jet's precession is a disc's 
intrinsic instability, which  warps the disc \citep{Pri97} and wobbles
the jet's direction on a time scale of hundreds of years \citep{Liv97}.
The two mechanisms predict similar precession time-scales, which are few times 
smaller  than a typical PN's age. This allows few precessions in 
total, enough   to produce the observed point-symmetric structure.
 
\begin{figure}[htb]
\begin{minipage}[t]{85mm}
\includegraphics[width =85mm, angle = -90]{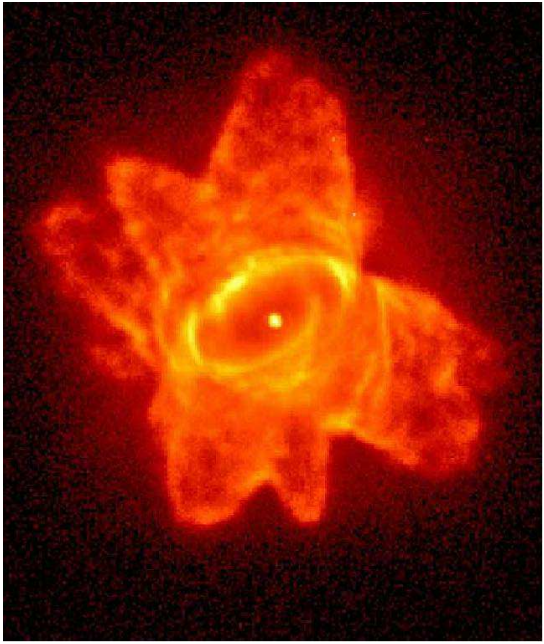}
\caption{\label{f:he247} The planetary nebula He~2-47 \citep{Sah00}; image 
by kind permission  of A.~R.~Hajian, http://ad.usno.navy.mil/pne/gallery.html.
The  angular size of this PN is about 9'' \citep{Sah00}.
}
\end{minipage}
\end{figure}

Even if both  mechanisms predict the same  time-scale for precession, 
there is a 
potential important difference between the two models, which will turn out to 
be important when we shall extend this model to galaxy clusters. 
The self-induced warping instability causes the accretion disc to
wobble in a stochastic  manner \citep{Pri97,Liv97}.
In such a case we expect that the signature will be a point-symmetric
nebula, but with no global axisymmetry.
Such nebulae might have a morphology resembling that  of the 
PN~He~2-47 (see Figure~\ref{f:he247}, from \citealp{Sah00}),
and to less extent that of the PN M1-37 \citep{Sah00}.
We point out that a model based on a stellar binary system at
the centre of each of these two PNe can also account for their structure.
In any case, we do not expect that the self-induced warping instability
will lead to a nice global axisymmetric structure ---~besides the
point-symmetric brightness~--- as in the PN~Hb~5 (Figure~\ref{f:hb5}).

\section{A Generalisation to Galaxy Clusters}
\label{s:clusters}

In this Section we   extend this physical picture to the X-ray 
cavities in  galaxy clusters. 
The last generation X-ray satellites {\sl Chandra} and {\sl XMM-Newton}
have revealed the presence of X-ray deficient bubble pairs
in the diffuse medium of several cooling flow clusters of galaxies.
(e.g. 
Hydra~A: \citet{McN00}; 
Perseus: \citet{Fab00,Fab03}; 
A~2597:  \citet{McN01}; 
A~4059:  \citet{Hei02}; 
RBS~797: \citet{Sch01};  
A~2052   \citet{Bla03}). 
These X-ray cavities also harbour a relevant radio emission, owing to 
synchrotron  radiation.
It is generally accepted that these cavities are associated to the
activity of an  AGN sitting  at the cluster centre. 
Its outbursts may inject a sizeable amount of energy ($10^{57}-10^{60}\erg$) 
into the diffuse intra-cluster medium (ICM). Magnetic fields and hot, 
possibly relativistic, plasma displace the dense ICM, leaving over thin and
X-ray faint cavities. The absence of strong shocks at the rims of essentially 
all these cavities shows that the ICM displacement is quite gentle 
\citep{McN04}.

In some cases the diffuse X-ray emission near these cavities  exhibits a
point-symmetric structure. An example is the recently observed galaxy 
cluster MS~0735.6+7421, whose {\sl Chandra} image 
was first presented at the 2004~{\sl COSPAR} meeting 
(\citealp{McN05}, see also Figure~\ref{f:ms}).
This is a faraway ($z\simeq 0.2$) cluster,
thus even with {\sl Chandra} its angular resolution is rather poor. 
To better appreciate the analogies  with   point-symmetric nebulae, we 
compare it to a
low-resolution image of the  point-symmetric nebula Hb~5
(Figure~\ref{f:hb5}, right panel). Both 
structures have similar point-symmetries: as discussed above, if we suppose 
they are due to a precessing jet, the regular morphology hints in both cases
for a binary-driven precession, rather than for a stochastic
precession {\`a} la  Pringle.   Even in this case, the morphological criterion
is the only discriminant, because the precession time scales predicted by the 
two models are similar:
\be
\tau_{\rm prec} \simeq 2\times 10^7 \yr \;\;
\alpha^{-1} \; \frac{M}{10^9\msun},
\ee
for Pringle's model (equation (4.11) of \cite{Pri97};  $\alpha$ is
the celebrated \citet{Sha73} alpha-parameter), and
\be
\tau_{\rm prec} \simeq  10^6 \yr \; 
\left(\frac{M}{10^9\msun}\right)^{1/2}
\left(\frac{a}{10^{19}\cm}\right)^3
\left(\frac{a_d}{10^{18}\cm}\right)^{-3/2}\; 
\frac{(1+q)^{1/2}}{q\: \cos\vartheta}.
\ee
for the binary model. For a typical mass ratio $q=0.1$, 
the precession period is  
$\tau_{\rm prec} \simeq 10^7 \yr$, close to the value 
predicted by the warp-instability model. Both the estimated
precession time scales are  few times  smaller than  the estimated lifetime 
of the jet activity $\simeq 10^8\yr$ \citep{McN04}, 
allowing few precession in total.
We conclude this Section with a {\em caveat}.  A precessing opposite-jet  
model may explain quite naturally  a point-symmetric structure like in  
MS~0735.6+7421, because it  can  influence symmetrically both  sides of the 
cluster's centre. Yet,  alternative explanations are by no way ruled out.
High resolution X-ray imaging, stellar dynamics analysis, detailed radio 
observations are  required  to  weed out  the spurious candidates.

\section{Summary and Conclusion}

In this paper we have been guided by the occurrence of point-symmetric
structures both in the X-ray deficient bubbles in some galaxy clusters
and in optical deficient bubbles in bipolar planetary nebulae.
We have pointed out that the ultimate origin of these symmetric
structures might be binary-driven precession of the jets
inflating the bubble pairs.
If this hypothesis is correct, the galaxy cluster MS~0735.6+7421
should harbour a massive binary black hole at its centre. 

We may look ahead beyond the special case of MS~0735.6+7421, and suggest
that a systematic cataloguing of point-symmetric X-ray structures
in galaxy clusters might increase the number of candidate MBBHs.
For this task a high angular resolution is necessary. {\sl Chandra} 
is suitable for that task, and in the future more candidates 
are  to be expected in the {\sl XEUS} sky.

\bigskip
The initial part of this research was motivated by discussions
with Mordechai Rorvig. 
We thank Brian McNamara  for useful comments.
We thank Arsen Hajian,
Raghvendra Sahai, and Romano Corradi for the images of the PNe.
FP is supported by a Fine Fellowship at the Technion-Israel
Institute of Technology.
FP was supported by grant No. 2002111 from the United
States-Israel Binational Foundation (BSF), Jerusalem, Israel.
This research was supported by the Israel Science Foundation.




\end{document}